\documentclass[preprint,aps,showpacs,amsfonts,epsf]{revtex4}

\input{epsf.tex}

\newcommand{\be}{\begin{equation}}
\newcommand{\ee}{\end{equation}}
\newcommand{\bea}{\begin{eqnarray}}
\newcommand{\eea}{\end{eqnarray}}
\newcommand{\ba}{\begin{array}}
\newcommand{\ea}{\end{array}}
\def\J#1#2#3#4{{#1} {\bf #2}, #3 (#4)}
\def\PRD{Phys. Rev. D}

\def\PTEP{Prog. Theor. Exp. Phys.}

\def\JMP{J. Math. Phys.}

\begin{document}

\title{Comment on ``Zero mass limit of Kerr spacetime is a
wormhole''}

\author{V. S. Manko}
\address{Departamento de F\'\i sica, Centro de Investigaci\'on y de
Estudios Avanzados del IPN, A.P. 14-740, 07000 Ciudad de M\'exico,
Mexico}

\begin{abstract}
It is argued that the main results of the paper of Gibbons and
Volkov (Phys. Rev. D {\bf 96}, 024053 (2017)) come from erroneous
interpretations of the surface $r=0$ in the Kerr and
Kerr--(anti)-de Sitter spacetimes.
\end{abstract}


\maketitle

In a recent paper of Gibbons and Volkov \cite{GVo} it has been
claimed that a specific configuration involving two copies of a
Minkowski space arises as the zero mass limit of the Kerr solution
\cite{Ker} extended to the negative values of the radial
coordinate $r$ in the manner of Boyer and Lindquist \cite{BLi}
(according to the book \cite{HEl}, for instance, the latter
extension has two asymptotic regions -- of positive and negative
mass -- connected on the disks $r=0$). A similar ``wormhole''
interpretation has also been given in \cite{GVo} to the zero mass
limit of the extended Kerr--(anti)-de Sitter solution \cite{Car}
whose asymptotic regions are glued by means of the disks $r=0$
too.

Apparently, the authors of \cite{GVo} overlooked some novel
results on the non-disk geometry of the surface $r=0$ in the Kerr
and Kerr--de Sitter spacetimes obtained and discussed recently in
the papers \cite{GMa,MGa,GMa2}. In the first of these papers the
surface $r=0$ of the Kerr solution was shown to be a dicone (see
Fig.~1), while in the second paper the surface $r=0$ of the
Kerr--(anti)-de Sitter solution was found to be a concave or a
convex dicone of constant Gaussian curvature $K=\Lambda/3$,
$\Lambda$ being a cosmological constant (see Fig.~2, and also
figures 2 and 3 of \cite{MGa}); in the third paper the specific
role of quasi-Cartesian coordinates in creating a misleading image
of the surface $r=0$ in the original papers on the analytical
extensions of the black-hole spacetimes was clarified. This shows
in particular that the maximal analytic extension of the Kerr
metric considered by Boyer and Lindquist in \cite{BLi}, which
treats the surface $r=0$ as a disk, is actually wrong. The correct
extension of the Kerr or Kerr--(anti)-de Sitter solution consists
in combining the interior region of the dicone with the main
manifold $r>0$, thus getting in each case a topologically trivial
spacetime with a single asymptotic region depicted in Fig.~3. Of
course, the zero mass limit of such extended solutions with one
asymptotic region will be a Minkowski or an (A)dS space,
respectively.

It follows from the above said that the extensions of the Kerr and
Kerr--(anti)-de Sitter spacetimes with two asymptotic regions of
positive and negative mass considered in the paper \cite{GVo}
actually do not exist as legitimate mathematical or physical
models, and hence there is no any sense in considering their zero
mass limit. It should be also noted that the analytical
cross-gluing on the disks considered in the paper \cite{GVo} is
{\it physically unrealizable}, being just a mathematical
abstraction, as the readers themselves can easily check for
instance by making identical cuts on two sheets of paper and then
trying to cross-glue them. At the same time, the abstract locally
Minkowskian (or (A)dS) spacetime of nontrivial topology with two
asymptotic regions certainly does exist as a purely mathematical
construction independently of the work \cite{GVo}, and everyone
can readily construct it as a mental experiment from two copies of
a Minkowski (or (A)dS) space. Moreover, using the results of the
paper \cite{GMa}, it is trivial to establish the true relation of
such spacetimes to the Kerr and Kerr--(anti)-de Sitter solutions.
Indeed, as was observed in \cite{GMa}, several identical copies of
the analytically extended Kerr spacetime (i.e., of the one shown
in Fig.~3) can give rise to the more sophisticated analytic
extensions of the Kerr solution characterized by several
asymptotic regions and nontrivial topology (such mathematical
constructions however will have a similar intrinsic ``gluing''
problem as the one arising during the construction of the Riemann
surface of the analytical function $\sqrt[n]{z}$, $n\ge2$). In the
case of two identical copies, after making the cuts in the
equatorial plane inside the singular rings (these cuts differ from
the $r=0$ surface!) and gluing mentally different sides of the
cuts of different copies, one readily gets a mathematical
construction whose zero mass limit is exactly the locally
Minkowskian space with two asymptotic regions considered in
\cite{GVo}. (The Kerr--(anti)-de Sitter case is treated
similarly.)

\section*{Acknowledgments}

The author is grateful to Eduardo Ruiz for interesting
discussions. This work was supported by the CONACyT of Mexico.




\begin{figure}[htb]
\centerline{\epsfysize=80mm\epsffile{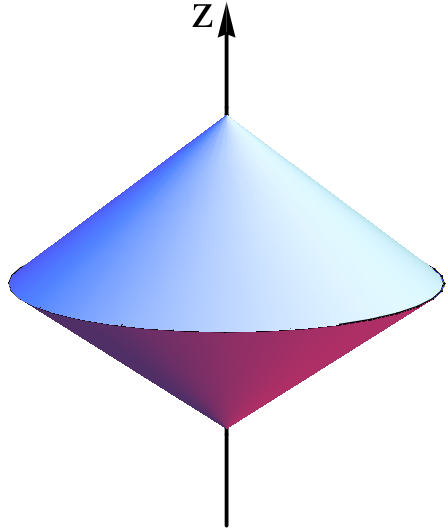}} \caption{The
surface $r=0$ in the Kerr solution -- a dicone.}
\end{figure}

\begin{figure}[htb]
\centerline{\epsfysize=80mm\epsffile{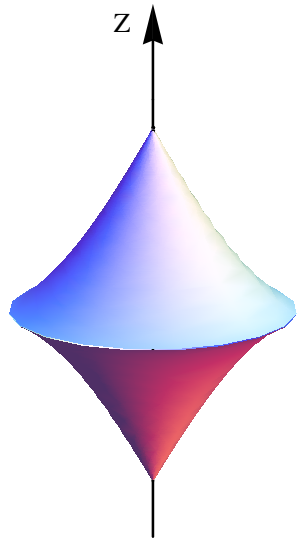}} \caption{The
surface $r=0$ in the Kerr--anti-de Sitter solution -- a concave
dicone of constant negative Gaussian curvature.}
\end{figure}

\begin{figure}[htb]
\centerline{\epsfysize=55mm\epsffile{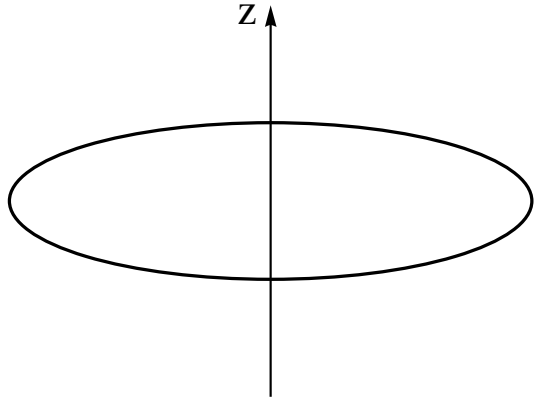}} \caption{The
analytically extended Kerr or Kerr--de Sitter solution with one
asymptotic region and traversable ring singularity.}
\end{figure}

\end{document}